\documentclass{iopart}
\usepackage{iopams}
\usepackage{graphicx}
\usepackage{cite}
\begin{document}
\title[The challenge of obtaining specific heat]
{Finite quantum dissipation: the challenge of obtaining specific heat}
\author{Peter H{\"a}nggi, Gert-Ludwig Ingold and Peter Talkner}
\address{Institut f{\"u}r Physik, Universit{\"a}t Augsburg, D-86135
Augsburg}
\ead{peter.hanggi@physik.uni-augsburg.de}
\begin{abstract}
We consider a free particle coupled with finite strength to a bath and
investigate the evaluation of its specific heat. A harmonic oscillator bath
of Drude type with cutoff frequency $\omega_{\mathrm{D}}$ is employed to
model an ohmic friction force with dissipation strength $\gamma$. Two
scenarios for obtaining specific heat are presented. The first one uses the
measurement of the kinetic energy of the free particle while the second one is
based on the reduced partition function. Both descriptions yield results
which are consistent with the Third Law of thermodynamics. Nevertheless,
the two methods produce different results that disagree even in their leading
quantum corrections at high temperatures. We also consider the regime where
the cutoff frequency is smaller than the friction strength, i.e.
$\omega_\mathrm{D}<\gamma$. There, we encounter puzzling results at low
temperatures where the specific heat based on the thermodynamic
prescription becomes negative. This anomaly is rooted in an ill-defined
density of states of the damped free particle which assumes unphysical
negative values when $\gamma/\omega_\mathrm{D}>1$. 
\end{abstract}
\submitto{\NJP}

\section{Introduction}
While some physical disciplines such as classical mechanics and
electrodynamics underwent profound changes with the birth of quantum
mechanics and relativity, thermodynamics proved impressively robust over
the last century. The main reason is that the formulation of thermodynamics
rests on few pillars only, such as entropy, temperature and the three Laws
relating these state variables. The grandness of thermodynamics is that
these concepts hold independently of the details of the corresponding total
system dynamics. Nonetheless, the statistical mechanical foundation of
thermodynamics strongly relies on the quantum mechanical properties of
matter in particular what concerns the low temperature behaviour.

Statistical mechanics gives rise to some subtle issues when going from
a closed description of all degrees of freedom, including those of
large environments, to a reduced description of an {\it open} system
where all bath degrees of freedom are traced over. Generally no
problems arise in the weak-coupling limit where the system-bath
interaction tends to zero. This is the situation typically assumed
explicitly, or at least implicitly, in the majority of textbooks. Even
in this limit, however, pitfalls can arise in the quantum case, as
recently elucidated in \cite{hanggi2005Chaos15}. Generally, care
must also be taken in defining correctly the expression for
(Gibbs) work in the First Law, as recent debates on the validity of
nonlinear fluctuation theorems have given evidence
\cite{jarzynski2007CRP8,horowitz2007JSM}.

In this work we emphasize the finiteness of the coupling between the open
quantum system and an environment of temperature $k_\mathrm{B}T =1/\beta$.
The finite coupling will be modelled in terms of a generalized quantum
Langevin equation (GLE) with a velocity-proportional memory friction, i.e.
ohmic damping. An equivalent microscopic approach is provided by a bilinear
coordinate-coordinate coupling between system degrees of freedom and
environmental degrees of freedom, see e.g. in \cite{hanggi2005Chaos15} and
references given therein. 

The presence of an ohmic friction term leads to an irreversible motion
with a unique stationary state. Already at this level, prominent
differences with the classical situation emerge. As is well known, the
classical canonical thermal equilibrium probability assumes the
familiar Gibbs-Boltzmann expression $\propto \exp(-\beta
H_\mathrm{S})$, with $H_\mathrm{S}$ being the bare system Hamiltonian
in absence of interaction. Most remarkably, this classical result
holds true independently of friction strength. In striking contrast,
the equilibrium density operator of an open quantum system becomes a
function of the friction strength \cite{hanggi2005Chaos15,grabe88},
thus exhibiting a dependence on the coupling to the environmental
degrees of freedom.  This feature can explicitly be inspected for an
ohmic-like, damped harmonic quantum oscillator
\cite{hanggi2005Chaos15,grabert1984ZPhysB55,riseborough1985PRA31}:
Only in the weak coupling limit does the canonical density operator
reduce to the common Gibbs state, where $H_\mathrm{S}$ then represents
the operator-valued system Hamiltonian.

Given this observation in the preceding paragraph it should not come as too
big a surprise that the evaluation of thermodynamic quantities for an open
quantum system, such, for example, as its specific heat, is also plagued
with subtleties. These difficulties all originate from the assumed finite
system-bath interaction. For the case of specific heat of a linear quantum
oscillator of finite friction strength this was explicitly demonstrated in
a recent work \cite{hangg06}. While in absence of damping zero entropy is
approached exponentially fast, this approach is weakened with finite
friction to a power law dependence in friction strength and temperature
\cite{hangg06,ford2007PRB75}. The Third Law has also been validated for the
quantum dissipative oscillator in the presence of velocity-coordinate and
velocity-velocity couplings \cite{wang08} as well as for a charged
oscillator in a magnetic field \cite{bandy08}. The main finding is that the
thermodynamic entropy of the open system vanishes according to a power law
in temperature with the same exponent that characterizes the frequency
dependence of the memory friction in the limit of vanishing frequency.

It was remarked in \cite{hangg06}, however, that the definition of the
specific heat is ambiguous and may lead to pronouncedly different values.
Therefore, the relation to experimental observations needs scrutiny, in
particular in the study of nanosystems that behave quantum in nature but
still are coupled with finite strength to an environment. Another
surprising observation is that the familiar von-Neumann entropy for a
quantum dissipative oscillator {\it fails} to approach zero for vanishing
bath temperature \cite{hoerh07}.

In the following we elucidate in detail the complications that arise
when evaluating thermodynamic quantities of open quantum systems whose
dissipation strength is finite. A most suitable test bed is the case
of a {\it free quantum particle} \cite{grabe88,hangg06}. It is our
working hypothesis that  this simple system better be understood first
before daring to embark on more complex physical situations. As is
well known, a classical free particle does not obey the Third Law.
Surprisingly, the coupling to a heat bath renders the system more
quantum thereby helping to restore the Third Law \cite{hangg06}.

The work is organized as follows. In the next section we couple a free
quantum particle to a heat bath which exerts a finite dissipation on
the free particle. We then focus on the evaluation of thermal
equilibrium quantities, in particular the specific heat. In
\sref{sec:tworoutes} we introduce two definitions of the specific heat
which a priori both seem physically well motivated. Their properties are
investigated in sections \ref{sec:energy} and \ref{sec:entropy}. The
emerging results cause worrisome ``baffling'' which we attempt to
resolve by inspecting more closely the underlying density of states of
a dissipative free particle. Some conclusions and consequences are
given in our final section.

\section{Coupling a free particle to a heat bath}\label{sec:principles}
\subsection{Dissipation  and corresponding quantum Langevin equation}
\begin{figure}
\includegraphics[width=0.6\textwidth]{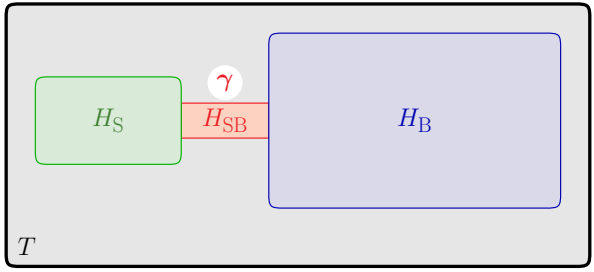}
\caption{Setup of an open nanosystem with Hamiltonian $H_\mathrm{S}$
which is coupled with finite strength $\gamma$ through an interaction
$H_\mathrm{SB}$ to a bath described by a Hamiltonian $H_\mathrm{B}$.
The total system composed of system plus bath is weakly coupled to a
super-bath which provides the working temperature $T$ in thermal
equilibrium.} 
\label{fig:hamil}
\end{figure}

As mentioned in the introduction a free particle, or more generally,
an ensemble of non-interacting, distinguishable particles fails to
approach zero entropy at accessible temperatures. This classical behaviour
is modified when the particle is able to exchange energy with a heat bath.
We therefore focus on the simplest situation of a free quantum particle 
of mass $M$ which is coupled to a heat bath made up of harmonic oscillators. 
The total Hamiltonian of particle plus bath thus reads (see \fref{fig:hamil})
\begin{equation}
\label{eq:def_H}
  H = H_\mathrm{S}+H_\mathrm{B}+H_\mathrm{SB}
\end{equation}
where
\begin{equation}
  H_\mathrm{S}=\frac{p^2}{2M}
\end{equation}
describes the free particle with momentum operator $p$, and
\begin{equation}
  H_\mathrm{B}=\sum_{i=1}^N\left(\frac{p_i^2}{2m_i}+\frac{m_i}{2}\omega_i^2
  x_i^2\right)
\end{equation}
represents a set of harmonic oscillators constituting the heat bath
which is bi-linearly coupled to the free particle via its position
operator $q$ by
\begin{equation}
  H_\mathrm{SB}=\sum_{i=1}^N\frac{m_i\omega_i^2}{2}\left(-2qx_i+q^2\right)\,.
\end{equation}
Here, the coupling constants have been chosen without loss of
generality in such a way that the ensemble of free particle and heat
bath is translationally invariant so that the damped particle can
still be considered as free.

Elimination of the heat bath leads to an effective equation of motion
for the position operator of the damped free particle, the quantum
Langevin equation of a free particle, reading
\begin{equation}
  \label{eq:eom}
  M \frac{\mathrm{d}^2}{\mathrm{d}t^2}q +  M \int_{t_0}^t\mathrm{d}s\gamma(t-s)
  \frac{\mathrm{d}}{\mathrm{d}s}q = \xi(t)\,.
\end{equation}
The coupling to the heat bath results in a damping kernel
\begin{equation}
  \gamma(t) =
  \frac{1}{M}\sum_{i=1}^N m_i\omega_i^2\cos(\omega_i t)
\end{equation}
and an operator-valued noise term
\begin{eqnarray}
\fl  \xi(t) = -M\gamma(t-t_0)q(t_0)\nonumber\\
  + \sum_{i=1}^N\left[m_i\omega_i^2x_i(t_0)
  \cos\big(\omega_i(t-t_0)\big)
  +\omega_ip_i(t_0)\sin\big(\omega_i(t-t_0)\right)
  \big] \;,
\end{eqnarray}
which depends on the initial conditions of free particle and heat
bath. The sum of the second and third contribution of this expression, i.e.
\begin{equation}
  \eta(t) = \xi(t)+M\gamma(t-t_0)q(t_0)\,,
\end{equation}
only depends on the initial positions and momenta of the heat bath and is 
characterized by a vanishing first moment,
\begin{equation}
  \langle\eta(t)\rangle_\mathrm{B}=0\,,
\end{equation}
while the symmetrized correlation function obeys
\begin{equation}
 \label{eq:etacorrelation}
\fl
\frac{1}{2}\langle\eta(t)\eta(s)+\eta(s)\eta(t)\rangle_\mathrm{B} =
  \frac{\hbar}{2}\sum_{i=1}^Nm_i\omega_i^3\cos\big(\omega_i(t-s)
  \big)\coth\left(\frac{\hbar\omega_i}{2k_\mathrm{B}T}\right) \,.
\end{equation}
Here, $\langle\dots\rangle_\mathrm{B}$ denotes an expectation value with
respect to the equilibrium density matrix 
$\exp(-\beta H_\mathrm{B})/\Tr[\exp(-\beta H_\mathrm{B})]$ of the isolated
bath. The temperature $T$ is imposed on the heat bath via the weak coupling
to a super-bath (see \fref{fig:hamil}). The commutator of the noise 
is non-vanishing, reading
\begin{equation}
  [\eta(t),\eta(s)]=-\mathrm{i}\hbar\sum_{i=1}^Nm_i\omega_i^3
  \sin\big(\omega_i(t-s)\big)\,,
\end{equation}
and guarantees that the familiar commutator-relation between position
$q$ and momentum $p$ is obeyed at all times, as required by quantum 
mechanics.

In the following, we will restrict ourselves to the so-called Drude
model of quantum dissipation where the damping kernel describes
exponential memory on the time scale $\omega_\mathrm{D}^{-1}$. For
positive arguments $t > 0$, the damping kernel assumes the form:
\begin{equation}
  \gamma(t)=\gamma\omega_\mathrm{D}\exp(-\omega_\mathrm{D}t)
\end{equation}
while  for negative times we formally obtain $\gamma(t<0) =
\gamma(|t|)$.  This in turn assures that the symmetrized correlation
function of the time-homogeneous noise correlation in
\eref{eq:etacorrelation} obeys a generalized Einstein relation
\cite{hanggi2005Chaos15}. The value $\gamma$ yields the damping
strength according to
\begin{equation}
  \int_0^{\infty}\mathrm{d}t\gamma(t) = \gamma\,.
\end{equation}
For later use we will also need the Laplace transform of the damping
kernel, which is given by
\begin{equation}
  \label{eq:gammadach}
  \hat\gamma(z) = \int_0^{\infty}\mathrm{d}t\gamma(t)\exp(-zt) =
  \frac{\gamma\omega_\mathrm{D}}{z+\omega_\mathrm{D}}\,.
\end{equation}

The total mass of the heat bath can be expressed via the appealing
formula \cite{grabe88}
\begin{equation}
  \sum_{i=1}^N m_i = M\lim_{z\rightarrow0}\frac{\hat\gamma(z)}{z}\,.
\end{equation}
As a consequence of the fact that the Drude model behaves ohmic for
low frequencies, i.e.\ $\hat\gamma(z)$ goes to a positive constant
$\gamma$ for $z\rightarrow0$, this heat bath behaves non-ballistic in
the sense that it assumes an infinite mass.

The Drude model is typically employed for regularization purposes. As
an example we mention that in the case of strict ohmic dissipation,
i.e.\ memoryless damping where $\gamma(t)=2\gamma\delta(t)$, the
second moment $\langle p^2\rangle$ of the momentum $p$ with respect to
the equilibrium density matrix of system plus bath exhibits an
ultraviolet divergence. For the purposes of the regularization of such
a quantity, it is assumed that $\omega_\mathrm{D}^{-1}$ is by far the
shortest time scale in the problem apart possibly from the thermal
time scale $\hbar\beta$. For a free particle subject to Drude damping
one therefore typically assumes $\gamma\ll\omega_\mathrm{D}$. In this
paper, we drop the requirement of a high-frequency cutoff which in
turn enables us to study also structured, uncommon environments
which may contain a low-frequency cutoff $\omega_\mathrm{D}$,  even
smaller than the damping strength $\gamma$.

The Drude model represents the simplest reservoir with memory in the
sense that the memory effect can be described by a single additional
degree of freedom. In fact, the deterministic equation of motion
\eref{eq:eom}, i.e. where $\xi(t)$ is set to zero, is equivalent to
\begin{eqnarray}
  \dot q &= v\nonumber\\
  \dot v &= z\nonumber\\
  \dot z &= -\omega_\mathrm{D}z-\gamma\omega_\mathrm{D}v\,.
  \label{eq:eoms}
\end{eqnarray}
This system of differential equations contains a zero-frequency mode
because system plus bath is translationally invariant. The remaining
two eigenfrequencies are obtained from the second and third equation
which describe a damped harmonic oscillator with damping strength
$\omega_\mathrm{D}$ and oscillator frequency
$(\gamma\omega_\mathrm{D})^{1/2}$. For $\omega_\mathrm{D}>4\gamma$,
one finds exponentially damped motion which for very large cutoff
frequencies contains the time scales $\gamma$ and $\omega_\mathrm{D}$
as expected. However, for sufficiently small cutoff frequency, where
$\omega_\mathrm{D} <4\gamma$, the eigenvalues become complex and one
observes a damped oscillation. This makes the Drude model more
interesting than one might initially expect.

\subsection{Thermodynamics of a dissipative free particle} 
We note that a thermodynamic description of a free particle can be
meaningful only if the particle is confined to stay within a box of finite
size. The presence of the box leads to the quantization of the energy
levels with the typical excitation energy  $\Delta E = \hbar^{2} \pi^{2}/2
M L^{2}$ between the first excited state and the ground state for a 
one-dimensional box of length $L$ with reflecting walls. This system in
isolation (i.e. in the absence of dissipation) approaches zero entropy only
at extremely low temperatures when the thermal energy is comparable to or
below the energy difference between the first excited and the ground state. 

We will restrict our investigations to temperatures which are high
enough such that the discreteness of the spectrum of a single free
particle can be ignored even in the weak coupling limit, i.e. we will
always assume that $\beta \Delta E \ll 1$. For a Helium atom at a
temperature of approximately 500\,pK -- the lowest temperature which
can be reached today in the laboratory -- the thermal energy and the
excitation energy are equal for a cavity of linear size of the order
of 30\,$\mu$m. Even for such extremely low temperatures a handy box of
linear size of, say, 1\,cm would suffice to meet the required
condition with $\beta \Delta E \approx 10^{-5}$.  In order that any
quantum effects can survive under thermal conditions rendering the
discreteness of the particle spectrum practically invisible, the bath
must provide relevant energy scales that are large compared to the
thermal energy. In the case of a bath with a Drude cutoff, these
relevant energies are given by the damping constant and the cutoff
frequency.  Hence, a regime of low temperatures with $\Delta E\ll
k_\mathrm{B}T\ll \hbar \gamma, \hbar \omega_\mathrm{D}$ exists where
quantum effects can be expected to become relevant.

\section{Free quantum Brownian motion: Two routes to calculate specific
heat}\label{sec:tworoutes}

We are interested in the specific heat of a free damped particle. The
volume in which the particle can move, will be assumed very large but
fixed. In this case, the specific heat is obtained from the internal
energy $U$ by taking the derivative with respect to the temperature
$T$, i.e.
\begin{equation}
  C=\frac{\partial U}{\partial T}\,.
\end{equation}

Usually, it is supposed that the coupling of the system to the heat
bath defining the temperature can be treated in the limit of vanishing
coupling strength. However, here we are interested in the case of
finite coupling where the meaning of the system's internal energy is
no longer obvious. In the following, we will study two different
approaches.

One possibility is to replace $U$ by the energy $E$ defined as
expectation value of the system Hamiltonian $H_\mathrm{S}$
\begin{equation}
  \label{eq:def_e}
  E= \langle H_\mathrm{S}\rangle\,,
\end{equation}
where 
\begin{equation}
\langle H_\mathrm{S}\rangle = \frac{\Tr_\mathrm{S+B}[H_\mathrm{S}
\exp(-\beta H)]}{\Tr_\mathrm{S+B}[\exp(-\beta H)]}\,.
\end{equation} 
This leads to our {\it first} definition of a specific heat
\begin{equation}
  \label{eq:ce}
  C^E = \frac{\partial E}{\partial T}\,.
\end{equation}
This definition is based on the system Hamiltonian $H_\mathrm{S}$
and includes the interaction of the system with the bath
only via the density matrix of the total system.

Alternatively, one can start from the well-known and widely used
expression for the partition function of the reduced system: It is
defined in terms of the partition functions of the coupled system and
of the uncoupled bath as
\cite{grabe88,grabert1984ZPhysB55,hangg06,ford2007PRB75,feynm63,feynman72,calde83,legge87,FLOC,hanke95,qtad98,ingol02}
\begin{equation}
  \label{eq:def_z}
  Z = \frac{\Tr_\mathrm{S+B}[\exp(-\beta H)]}
  {\Tr_\mathrm{B}[\exp(-\beta H_\mathrm{B})]} \;,
\end{equation}
where the total Hamiltonian $H$ consists of contributions from the system,
the bath, and the coupling according to \eref{eq:def_H}. Employing the
standard relation between partition function and internal energy
\begin{equation}
  \label{eq:def_u}
  U = -\frac{\partial}{\partial\beta}\mathrm{ln}(Z)\,,
\end{equation}
\eref{eq:def_z} implies that the internal energy is defined as
\begin{eqnarray}
\label{UvsE}
  U &= \langle H\rangle-
                        \langle H_\mathrm{B}\rangle_\mathrm{B}\nonumber\\
                     &=  E + [\langle H_\mathrm{SB}\rangle
                         +\langle H_\mathrm{B}\rangle
                         -\langle H_\mathrm{B}\rangle_\mathrm{B}]\,.
\end{eqnarray}
While the internal energy $U$ and the system energy $E$ agree in the
absence of a coupling between system and bath, this is no longer the
case at finite coupling. Particularly disturbing is the observation
that this difference is not solely given by the expectation value of the
interaction Hamiltonian $H_\mathrm{SB}$, but still contains the
difference of the bath energies caused by the interaction with the 
system.

We can now proceed and define a {\it second} specific heat, reading
\begin{equation}
  \label{eq:cz}
  C^Z = \frac{\partial U}{\partial T} \;.
\end{equation}
The partition function \eref{eq:def_z} also allows us to define an
entropy
\begin{equation}
  \label{eq:s}
  S=k_\mathrm{B}\left[\mathrm{ln}(Z)-\beta\frac{\partial}{\partial\beta}
                      \mathrm{ln}{Z}\right]\,.
\end{equation}
This results from the thermodynamic relation
\begin{equation}
  S=-\frac{\partial F}{\partial T}
\end{equation}
with the free energy $F=-(1/\beta)\mathrm{ln}(Z)$.

Alternatively, the second specific heat \eref{eq:cz} can as well be
derived from the entropy \eref{eq:s}; i.e. $C^Z$ is equivalently
obtained as
\begin{equation}
  \label{eq:czfroms}
  C^Z = T\frac{\partial S}{\partial T}\,.
\end{equation}

Our main concern is that these two routes of obtaining a specific heat
generally yield different results. This will be elucidated further by
studying the dissipation model of the Drude form when applied
to the simple quantum dynamics of a free particle.

\section{Follow the route using energy}\label{sec:energy}
We start our exploration of the specific heat by employing the
definition \eref{eq:ce} based on the expectation value of the system
energy
\begin{equation}
  E = \frac{1}{2M}\langle p^2\rangle\,.
\end{equation}
Here, $M$ denotes the mass of the particle and $p$ the momentum
operator. For a general heat bath leading to an equation of motion of
the form \eref{eq:eom} and for thermal energies $\beta^{-1}$ which are
large compared to the energy scale $\hbar^{2}\pi^{2}/2ML^{2}$ one
finds \cite{grabe88}
\begin{equation}
  \label{eq:edamped}
  E=\frac{1}{2\beta}\left[1+2\sum_{n=1}^\infty\frac{\hat\gamma(\nu_n)}
                          {\nu_n+\hat\gamma(\nu_n)}\right]
\end{equation}
with the Matsubara frequencies $\nu_n=2\pi n/\hbar\beta$. In the
high-temperature limit, the sum vanishes and one recovers
$E=k_\mathrm{B}T/2$, in agreement with the classical equipartition
theorem. For strictly ohmic damping, $\hat\gamma(z)=\gamma$, the sum
diverges as mentioned above and one is obliged to introduce a cutoff
in the spectral density of the heat bath. From \eref{eq:edamped} we 
obtain by means of \eref{eq:ce} the specific heat
\begin{equation}
  \label{eq:cesum}
  \frac{C^E}{k_\mathrm{B}} = \frac{1}{2}+\sum_{n=1}^{\infty}
  \frac{\hat\gamma^2(\nu_n)+\nu_n^2\hat\gamma'(\nu_n)}
       {(\nu_n+\hat\gamma(\nu_n))^2}\,,
\end{equation}
where the prime denotes a derivative with respect to the argument. 

For strictly ohmic damping one has $\hat\gamma(z)=\gamma$ and therefore
$\hat\gamma'(z)=0$. It is tempting to drop the second term in the numerator
of the sum and to evaluate the resulting sum which converges even for
strict ohmic damping. However, this would yield a specific heat which
diverges as the temperature approaches zero (see also the discussion after
\eref{eq:ceohmic}). The reason for this unphysical behaviour is that
for constant $\hat\gamma(z)$ the sum in \eref{eq:edamped} does not converge
and therefore derivatives should not be taken term-by-term.

\begin{figure}
\includegraphics[width=0.6\textwidth]{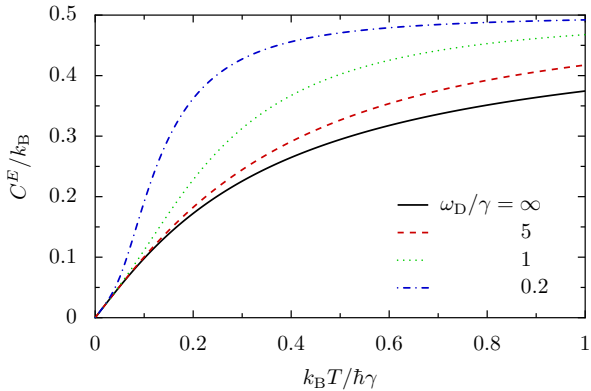}
\caption{Energy route: The specific heat $C^{E}$ is depicted as a
function of dimensionless temperature for different ratios of
$\omega_\mathrm{D}/\gamma$. The approach towards the classical limit $C^{E} =
0.5$ is at low temperatures inversely proportional to friction
strength $\gamma$; it notably becomes enhanced for decreasing values
of cutoff frequencies $\omega_\mathrm{D}$. The specific heat itself is always
positive and monotonically increases with increasing
temperature $T$.} \label{fig:Drude_E}
\end{figure}

For the Drude model, one obtains with \eref{eq:gammadach} for the
specific heat $C^{E}$:
\begin{equation}
  \label{eq:cedrude}
  \frac{C^E}{k_\mathrm{B}} = \frac{x_1x_2}{x_1-x_2}\left[x_2\psi'(x_2)
                               -x_1\psi'(x_1)\right]-\frac{1}{2} \;,
\end{equation}
where $\psi'(x)$ denotes the trigamma function and
\begin{equation}
  \label{eq:x12}
  x_{1,2}= \frac{\hbar\beta\omega_\mathrm{D}}{4\pi}\left(1\pm
                 \sqrt{1-\frac{4\gamma}{\omega_\mathrm{D}}}\right)\,.
\end{equation}
The behaviour of this result is depicted in \fref{fig:Drude_E} for
differently structured ohmic baths  as characterized by various ratios
of cutoff $\omega_\mathrm{D}$ over damping strength $\gamma$.

We next study some asymptotic limits more closely. For {\it high}
temperatures $T$ much larger than $\hbar\gamma/k_\mathrm{B}$ and
$\hbar\omega_\mathrm{D}/k_\mathrm{B}$, the specific heat approaches
the expected classical result with the leading quantum corrections
reading
\begin{equation}
  \label{eq:cehight}
  \frac{C^E}{k_\mathrm{B}}=\frac{1}{2}-\frac{\hbar^2\gamma\omega_\mathrm{D}}
  {24(k_\mathrm{B}T)^2}+\Or(T^{-3})\,.
\end{equation}
At {\it low} temperatures, i.e for $k_{\mathrm{B}}T \ll \hbar
\gamma,\; \hbar \omega_\mathrm{D}$  but still
$k_{\mathrm{B}}T\gg\hbar^{2}\pi^{2}/2ML^{2}$ the specific heat goes to
zero linearly according to
\begin{equation}
  \label{eq:celowt}
  \frac{C^E}{k_\mathrm{B}} = \frac{\pi}{3}\frac{k_\mathrm{B}T}{\hbar\gamma}
   -\frac{4\pi^3}{15}\left(\frac{k_\mathrm{B}T}{\hbar\gamma}\right)^3
   \left(1-2\frac{\gamma}{\omega_\mathrm{D}}\right)+\Or(T^5)\,.
\end{equation}

This behaviour is in agreement with the Third Law: Finite quantum
dissipation thus restores the validity of the Third Law of a free particle
already at temperatures for which the discreteness of the particle spectrum
is still irrelevant. The limiting behaviour is inversely proportional to
damping strength. Put differently, strong dissipation diminishes the
prefactor and thereby yields a turnover to the classical behaviour at higher
temperatures only. The leading low-temperature behaviour of the specific
heat only depends on the low-frequency behaviour of $\hat\gamma(z)$.
Notably, only the next-to-leading order also depends on the cutoff
frequency $\omega_\mathrm{D}$.

In the limit of an infinite cutoff frequency
$\omega_\mathrm{D}\rightarrow\infty$  the complete temperature
dependence  is obtained from \eref{eq:cedrude} as
\begin{equation}
  \label{eq:ceohmic}
  \frac{C^E}{k_\mathrm{B}} = \left(\frac{\hbar\gamma}
                                    {2\pi k_\mathrm{B}T}\right)^2
                  \psi'\left(\frac{\hbar\gamma}{2\pi k_\mathrm{B}T}\right)
                  -\frac{\hbar\gamma}{2\pi k_\mathrm{B}T}-\frac{1}{2}\,.
\end{equation}
The second term would have been missed in a naive calculation where
one simply sets $\gamma'(z)=0$ and subsequently evaluates the sum in
\eref{eq:cesum}. This term is important because it  ensures that the
specific heat does not diverge for $T\rightarrow0$.

\section{Follow the route using the thermodynamic partition
function}\label{sec:entropy}
In order to obtain the specific heat from the partition function
\eref{eq:def_z}, one must first determine the latter quantity for the
damped free particle. 

We again assume that the volume available to the particle is sufficiently
large so that the discreteness of the energy levels can be neglected. In
the limit of vanishing coupling to the heat bath, the partition function is
given by
\begin{equation}
  Z_0 = \frac{L}{\hbar}\left(\frac{2\pi m}{\beta}\right)^{1/2} \;.
\end{equation}
This result is a consequence of the density of states $\rho(E)\sim
E^{-1/2}$ of a free particle in one dimension.

In the presence of a finite coupling $\gamma$ to the heat bath, the
partition function is modified by the ratio of the two fluctuation
determinants for the free particle in the absence and presence of the
dissipative coupling \cite{grabe88,feynm63,feynman72,FLOC}.
We then obtain for this so defined  partition function
\begin{equation}
  \label{eq:z}
Z = \frac{L}{\hbar}\left(\frac{2\pi m}{\beta}\right)^{1/2}\prod_{n=1}^\infty
\frac{\nu_n}{\nu_n+\hat\gamma(\nu_n)}
\end{equation}
which will represent the starting point for all following calculations
discussed in this section. We recall that the modifications due to the
environmental coupling do not take into account the presence of
confining walls. In using this expression for the partition function,
we therefore assume a sufficiently large confining box so that
boundary effects, which could show up in the partition function, can
safely be neglected.

The specific heat can be calculated from the partition function either
by means of the internal energy \eref{eq:cz} or the entropy
\eref{eq:czfroms}. Both routes lead to identical results. In order to
stay close to the reasoning of the previous section, we choose the
first alternative.

The internal energy of a damped free particle is obtained from
\eref{eq:z} by means of \eref{eq:def_u}, reading
\begin{equation}
\label{eq:internalenergy}
U = \frac{1}{2\beta}\left[1+2\sum_{n=1}^\infty\frac{\hat\gamma(\nu_n)-
\nu_n\hat\gamma'(\nu_n)}{\nu_n+\hat\gamma(\nu_n)}\right]\,.
\end{equation}
This internal energy differs from the energy \eref{eq:edamped} by the
second term of the numerator appearing in the sum.

For the Drude model, the sum can be evaluated and the internal energy
is obtained as
\begin{equation}
  \label{eq:udrude}
  U = \frac{\hbar\omega_\mathrm{D}}{2\pi}\psi\left(\frac{\hbar\beta
  \omega_\mathrm{D}}{2\pi}\right)-\frac{x_1}{\beta}\psi(x_1)
  -\frac{x_2}{\beta}\psi(x_2)-\frac{1}{2\beta} \;,
\end{equation}
where $\psi(x)$ is the digamma function and $x_{1,2}$ are defined in
\eref{eq:x12}. This result should be contrasted with the energy $E$
for which one finds from \eref{eq:edamped} for the Drude model
\begin{equation}
  E = \frac{x_1x_2}{\beta(x_1-x_2)}\left(\psi(x_1)-\psi(x_2)\right)
      -\frac{1}{2\beta}\,.
\end{equation}
For $\hbar\beta\omega_\mathrm{D}/2\pi\gg1$, the internal energy $U$
and the energy $E$ thus differ by a constant energy contribution
which does not play a role for the specific heat,
but nevertheless indicates that the two quantities do not agree. This
disagreement even persists  in the case of strict ohmic damping. Note,
however, that in this limit both quantities diverge logarithmically in
the cutoff frequency $\omega_\mathrm{D}$, so that $E$ and $U$ will
contain an infinite energy contribution independent of temperature.

\begin{figure}
\includegraphics[width=0.6\textwidth]{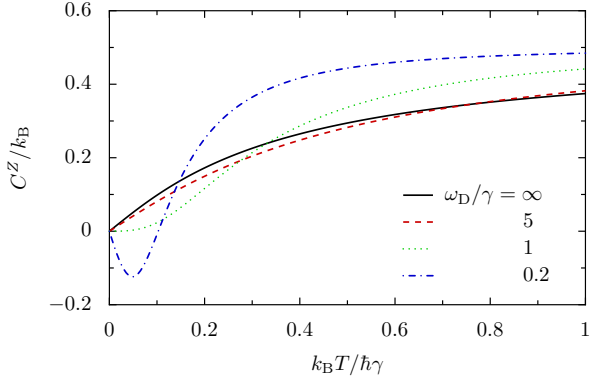}
\caption{Partition function route: The specific heat $C^{Z}$ is
depicted as a function of dimensionless temperature for different  
ratios of $\omega_\mathrm{D}/\gamma$. In the strict ohmic limit
$\omega_\mathrm{D}\rightarrow\infty$ this result agrees with $C^{E}$,
as depicted in \fref{fig:Drude_E}. The behaviour as a function of
finite cutoff is intriguing: While for $\omega_\mathrm{D}> \gamma$,
$C^{Z}$ is monotonically growing with increasing temperature, as
expected from thermodynamics, this fails to be the case  when the
friction strength surpasses the cutoff strength. Although still
vanishing at absolute zero temperature we find at low temperatures
formally negative-valued specific heats for this dissipative
quantum system.} 
\label{Drude_Z}
\end{figure}

From \eref{eq:udrude} it is  straightforward to evaluate the specific
heat $C^{Z}$  by means of \eref{eq:cz}; i.e.,
\begin{equation}
\label{C_z}
  \frac{C^Z}{k_\mathrm{B}} = x_1^2\psi'(x_1)+x_2^2\psi'(x_2)
  -\left(\frac{\hbar\beta\omega_\mathrm{D}}{2\pi}\right)^2
  \psi'\left(\frac{\hbar\beta\omega_\mathrm{D}}{2\pi}\right)-\frac{1}{2}\,.
\end{equation}
In the strictly ohmic limit $\omega_\mathrm{D}\to\infty$, this result
agrees with the corresponding expression \eref{eq:ceohmic} for $C^E$.
Some results obtained from \eref{C_z} are depicted in \fref{Drude_Z}.

For {\it high} temperatures but finite cutoff, i.e. if
$\hbar\beta\gamma/2\pi, \hbar\beta\omega_\mathrm{D}/2\pi\ll1$, one
obtains
\begin{equation}
\label{eq:CZ}
  \frac{C^Z}{k_\mathrm{B}} = \frac{1}{2}-\frac{\hbar^2\gamma\omega_\mathrm{D}}
  {12(k_\mathrm{B}T)^2}+\Or(T^{-3})\,.
\end{equation}
While this reproduces the correct classical result, the leading
correction differs by a factor of two from the high-temperature result
of the specific heat \eref{eq:cehight} derived from the system energy
$E$.

Very puzzling is the behaviour at {\it low} temperatures.  In this
limit  the specific heat behaves like
\begin{equation}
  \label{eq:czlowt}
  \fl
  \frac{C^Z}{k_\mathrm{B}} = \frac{\pi}{3}\frac{k_\mathrm{B}T}{\hbar\gamma}
  \left(1-\frac{\gamma}{\omega_\mathrm{D}}\right) -\frac{4\pi^3}{15}
  \left(\frac{k_\mathrm{B}T}{\hbar\gamma}\right)^3
  \left[1-3\frac{\gamma}{\omega_\mathrm{D}}-
  \left(\frac{\gamma}{\omega_\mathrm{D}}\right)^3\right]+\Or(T^5)\,.
\end{equation}
Notably this result  does again obey the Third Law in the same sense
as $C^{\mathrm{E}}$ does. Moreover, \eref{eq:czlowt} agrees with
\eref{eq:celowt} for $\omega_\mathrm{D}\gg \gamma$.  The big surprise
surfaces, however, for small cutoff frequencies. For
$\omega_\mathrm{D}<\gamma$, the specific heat $C^{Z}$ 
becomes negative at very low temperatures. This indicates a
serious defect which must be hidden somewhere in the theory leading to
\eref{eq:czlowt}.

The appearance of a negative specific heat is not restricted to the
Drude model. This can be seen by expanding the general expression
\eref{eq:internalenergy} for the internal energy in powers of
temperature. By means of the Euler-MacLaurin summation formula one
obtains
\begin{equation}
\label{eq:uexpansion}
\fl U = U_0 + \sum_{n=1}^\infty\frac{B_{2n}}{(2n)!}
\left(\frac{2\pi}{\hbar}\right)^{2n-1}(k_\mathrm{B}T)^{2n}\left[
f^{(2n-1)}(\infty)-f^{(2n-1)}(0)\right]
\end{equation}
where $B_{2n}$ are the Bernoulli numbers, $f^{(k)}$ denotes the $k$-th
derivative of $f$ and the ground state energy is given by
\begin{equation}
U_0 = \frac{\hbar}{2\pi}\int_0^\infty\mathrm{d}xf(x)\,.
\end{equation}
For the free damped particle, one finds by means of \eref{eq:internalenergy}
\begin{equation}
f(x) =\frac{\hat\gamma(x)-x\hat\gamma'(x)}{x+\hat\gamma(x)}\,.
\end{equation}
From \eref{eq:uexpansion} one readily obtains the specific heat
\begin{equation}
\frac{C^Z}{k_\mathrm{B}} = \sum_{n=1}^\infty\frac{B_{2n}}{(2n-1)!}
\left(\frac{2\pi k_\mathrm{B}T}{\hbar}\right)^{2n-1}\left[
f^{(2n-1)}(\infty)-f^{(2n-1)}(0)\right]\,.
\end{equation}
The leading term of this expansion yields
\begin{equation}
\frac{C^Z}{k_\mathrm{B}}=\frac{\pi}{3}\frac{1+\hat\gamma'(0)}{\hat\gamma(0)}
\frac{k_\mathrm{B}T}{\hbar}+\Or(T^3)
\end{equation}
in agreement with the specific heat \eref{eq:czlowt} for the Drude
model. This result implies that a negative specific heat will be found
within the approach based on the partition function for every damping
kernel with $\hat\gamma'(0)<-1$.

In contrast, for a damped harmonic oscillator with frequency
$\omega_0$, one obtains
\begin{equation}
f(x) = \frac{2\omega_0^2+x\hat\gamma(x)-x^2\hat\gamma'(x)}
{\omega_0^2+x\hat\gamma(x)+x^2}
\end{equation}
resulting in the leading low-temperature behaviour of the specific
heat
\begin{equation}
\frac{C^Z}{k_\mathrm{B}}=\frac{\pi}{3}\frac{\hat\gamma(0)}{\omega_0^2}
\frac{k_\mathrm{B}T}{\hbar}+\Or(T^3)\,.
\end{equation}
Here, the specific heat remains positive independently of the damping
kernel.

\section{Partition function and density of states}
The theory leading to the unexpected appearance of a negative specific
heat, cf.\ \eref{eq:czlowt}, is based on an assumption and an
approximation. The assumption is that the thermodynamic behaviour can
be described by the partition function \eref{eq:def_z}. The
approximation concerns the calculation of this partition function
and limits the validity of the specific heat
\eref{eq:CZ} to sufficiently high temperatures for a fixed size of the
box confining the particle. By choosing a sufficiently large box one
though can approach such low temperatures for which the specific heat
$C^{\mathrm{E}}$ becomes negative but for which the approximation
still is fully justified.

It is therefore the form of the partition function as the ratio of the
two partition functions of the total system and the bath which
deserves further scrutiny. We will analyze the situation within the
Drude model. Inserting the Laplace transform \eref{eq:gammadach} of the
damping kernel into the general expression \eref{eq:z} of the
partition function, one finds
\begin{equation}
  Z=\frac{L}{\hbar}\left(\frac{2\pi m}{\beta}\right)^{1/2}
  \frac{\Gamma(1+x_1)\Gamma(1+x_2)}
  {\Gamma(1+\hbar\beta\omega_\mathrm{D}/2\pi)}\,,
\end{equation}
where $\Gamma(z)$ is the gamma function and $x_{1,2}$ are defined in
\eref{eq:x12}.

Formally, the partition function can be related to a density of states
$\rho(E)$ of a damped system by means of a Laplace transform
\cite{hanke95}
\begin{equation}
  \label{eq:z_rho}
  Z(\beta) = \int_0^\infty\mathrm{d}E \rho(E) \exp(-\beta E)\,.
\end{equation}
By its very definition as the number of states at energy $E$ per unit
energy the density of states $\rho(E)$ must not assume negative
values. This basic property of the density of states also restricts
the admissible form of physically meaningful partition functions. In
the remainder of this section we demonstrate that the unphysical
behaviour of the specific heat as given by \eref{eq:czlowt} comes
along with negative regions of the density of states.

In order to discuss the partition function and the density of states
it is useful to shift the origin of the energy scale to the ground
state energy of the damped free particle
\begin{equation}
  U_0=\frac{\hbar\omega_1}{2\pi}\mathrm{ln}\left(\frac{\omega_\mathrm{D}}
  {\omega_1}\right)
  +\frac{\hbar\omega_2}{2\pi}\mathrm{ln}\left(\frac{\omega_\mathrm{D}}
  {\omega_2}\right)\,.
\end{equation}
This expression is obtained from \eref{eq:udrude} in the limit of zero
temperature. Instead of the partition function $Z$ we thus consider
$Z\exp(\beta U_0)$.

\begin{figure}
\includegraphics[width=0.6\textwidth]{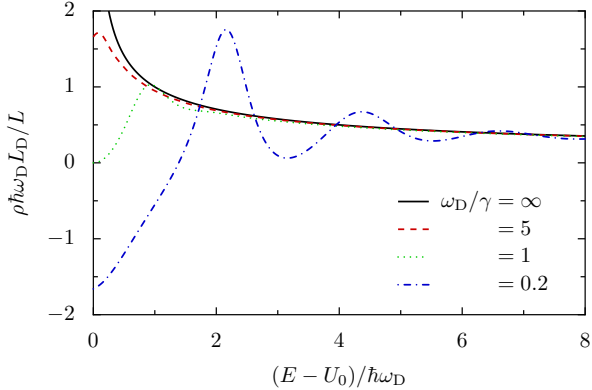}
\caption{Energy dependence of the density of states for 
$\omega_\mathrm{D}/\gamma=0.2, 1,$ and 5.
The solid line corresponds to the density of states of an undamped
particle in one dimension.}
\label{fig:rho}
\end{figure}

As the specific heat \eref{eq:czlowt} may become negative at low
temperatures, we focus on the behaviour of the density of states at
low energies. For $\hbar\beta\omega_\mathrm{D}\gg 1$ one finds
\begin{eqnarray}
\label{eq:z_shifted}
\fl Z\exp(\beta U_0) = \frac{L}{L_\mathrm{D}}\left(\frac{\pi\gamma}
{\omega_\mathrm{D}}\right)^{1/2}\nonumber\\\times\left[1+
\frac{\pi}{6\hbar\beta\omega_\mathrm{D}}\left(\frac{\omega_\mathrm{D}}{\gamma}-1\right)
+\frac{\pi^2}{72(\hbar\beta\omega_\mathrm{D})^2}\left(\frac{\omega_\mathrm{D}}{\gamma}-1\right)^2
+\Or(\beta^{-3})\right]
\end{eqnarray}
where
\begin{equation}
L_\mathrm{D} = \left(\frac{\hbar}{2m\omega_\mathrm{D}}\right)^{1/2}
\end{equation}
is a characteristic length related to the Drude cutoff frequency
$\omega_\mathrm{D}$. The inverse Laplace transform of
\eref{eq:z_shifted} yields the low-energy behaviour of the density of
states
\begin{eqnarray}
\fl\rho(E) = \frac{L}{\hbar\omega_\mathrm{D}L_\mathrm{D}}\left(\frac{\pi\gamma}
{\omega_\mathrm{D}}\right)^{1/2}\Bigg[\delta\left(\frac{E-U_0}
{\hbar\omega_\mathrm{D}}\right)\nonumber\\
+\frac{\pi}{6}\left(
\frac{\omega_\mathrm{D}}{\gamma}-1\right)+\frac{\pi^2}{72}\left(
\frac{\omega_\mathrm{D}}{\gamma}-1\right)^2\frac{E-U_0}{\hbar\omega_\mathrm{D}}
+\Or\left(\left(\frac{E-U_0}{\hbar\omega_\mathrm{D}}\right)^2\right)\Bigg]\,.
\end{eqnarray}
Except for the delta function at the ground state energy $U_0$, the
low-energy behaviour of the density of states is dominated by a
constant which changes sign at $\gamma=\omega_\mathrm{D}$. If the
cutoff frequency $\omega_\mathrm{D}$ is smaller than the damping
constant $\gamma$, the density of states is no longer positive
everywhere. This readily explains the surprising behaviour of the
specific heat found in the previous section. In \fref{fig:rho}, the
energy dependence of the density of states is shown for
$\omega_\mathrm{D}/\gamma = 0.2, 1,$ and 5. For
$\omega_\mathrm{D}<\gamma$, the density of states starts out with
negative values as expected. At larger energies, peaks of the density
of states are observed which can be related to the characteristic 
frequencies of the equation of motion
\eref{eq:eoms} which, for $\omega_\mathrm{D}<\gamma$ give rise to
underdamped oscillations. For sufficiently large damping constant
$\gamma$, there may exist several energy regions displaying a negative
density of states. For $\omega_\mathrm{D} \gg\gamma$, one recovers the
limit of vanishing damping where
\begin{equation}
\rho_0(E) = \frac{L}{L_\mathrm{D}}\frac{1}{(\hbar\omega_\mathrm{D}E)^{1/2}}\,.
\end{equation}

In the previous section, we have seen that the specific heat of a
damped harmonic oscillator at low temperatures is always positive. An
analysis along the lines presented above for the free damped particle
indeed shows that the density of states, apart from a delta-function
contribution, starts out at low temperatures with a positive value.
However, the positivity condition of a density of states defined by
\eref{eq:z_rho} is by no means obvious as we will discuss next.

By its statistical mechanical definition  the partition function is
the sum over the Boltzmann factors counted with the multiplicity of
the respective eigenenergies and therefore coincides with the Laplace
transform of the density of states as a function of the energy.  The
corresponding Laplace variable is the inverse temperature $\beta$.
This in particular implies that the inverse Laplace transform of any
proper partition function must not assume any negative values.

Apparently, the positivity of the density of states is not
automatically guaranteed by the above definition \eref{eq:def_z} as
can be seen in the case of a free particle. Another example is
provided by a system which is coupled to a rather trivial bath
consisting of a single oscillator with angular frequency $\omega$. In
this case one obtains for the partition function of the reduced
system, according to the definition \eref{eq:def_z}
\begin{equation}
\label{eq:z_so}
Z=\frac{\Tr_\mathrm{S+osc}[\exp(-\beta H)]}{\Tr_\mathrm{osc}[\exp(\-\beta
H_\mathrm{osc})]}=\sum_n g_n\mathrm{e}^{-\beta E_n}
\left(\mathrm{e}^{\hbar\beta\omega/2}-\mathrm{e}^{-\hbar\beta\omega/2}\right)
\end{equation}
where $E_{n}$ denotes the eigenenergies of the system coupled to the
oscillator and the positive weights $g_{n}$ are the corresponding 
degeneracies. For
a finite, spatially confined system, these eigenenergies constitute a
discrete spectrum. The first factor of the right-hand side coincides
with  the partition function of the total system, $Z_{\mathrm{S+B}}$
and the second term represents the inverse of the partition function of the
bath consisting of a single oscillator. The inverse Laplace transform
$\rho(E)$ of $Z$ becomes
\begin{equation}
\label{eq:rho}
\rho(E) = \sum_{n}g_{n} \delta(E-E_{n} + \hbar \omega/2) - \sum_{n}g_{n}
\delta(E-E_{n}-\hbar \omega/2)
\end{equation}
It consists of two sets of delta functions, one with positive weights
at the energies $E_{n}-\hbar \omega/2$ whereas the second set, which is
located at the shifted energies $E_{n} +\hbar \omega/2$, possesses
negative weights. Therefore, neither does $\rho(E)$ represent a
physically meaningful density of states nor must its Laplace transform
$Z= \int \mathrm{d}E\exp(-\beta E)\rho(E)$ be interpreted as the
partition function of the reduced system.

For reservoirs which represent true thermal baths both the total
system and the reservoir alone will have a continuous spectrum such
that a more subtle compensation of the different contributions to the
inverse Laplace transform may indeed lead to a positive density of
states and a corresponding, physically meaningful partition function.

\section{Conclusions}
In this work we compared two definitions of specific heat for a free
quantum particle which interacts at a finite strength with an oscillator
heat bath possessing a Drude cutoff. The first definition was based on
the assumption that the equilibrium expectation value of the
Hamiltonian $H_{\mathrm{S}}$ of the system can be identified with the
thermodynamic internal energy of the particle. The resulting specific
heat approaches the classical result for high temperatures and
goes to zero for vanishing temperature in accordance with the Third
Law.  The second definition is based on the ratio of the partition functions
of the total system and the heat bath which traditionally is
identified as the partition function of an open quantum system
\cite{grabe88,grabert1984ZPhysB55,hangg06,ford2007PRB75,feynm63,feynman72,calde83,legge87,FLOC,hanke95,qtad98,ingol02}.

The two definitions yield different results which even disagree
in the high temperature regime in terms of their quantum
corrections. If one associates (as done commonly) to each of the three
partition functions of the total system, the system and the bath a
free energy, the free energy of the open system is
expressed as the difference of the free energies of the total system
and the heat bath, in accordance with the ``most remarkable formula''
of Ford, Lewis and O'Connell \cite{FLOC}. We found that this
definition may lead to unphysical results for the damped free
particle, such as negative specific heat and negative regions of the
density of states. A corresponding analysis of the respective
quantities for a damped harmonic quantum oscillator though does not show
any violations of their formal properties
\cite{hangg06,ford2007PRB75}. This latter fact alone  does not resolve
the encountered baffling: The finding of a thermodynamic consistent
result which is in no apparent violation with formal properties, such
as a truly positive-valued density of states, does not necessarily
imply that the so obtained values are physically meaningful.

The definition of the specific heat via the expectation value of the
system Hamiltonian $H_{\mathrm{S}}$ leads to a quantity that can be
measured in principle. In contrast, no clear scheme seems to be
available for a measurement of the specific heat that derives from the
partition function of the reduced system. 

The conundrum of the physical meaning of a partition function of a reduced
system that is coupled with finite strength to an environment thus still
remains.  We emphasize once more that the detected difficulties all
``evaporate'' when the dissipation can be described within the
weak-coupling limit.  However, the impressive experimental progress 
on tailored, well-defined nanosystems that behave quantum in
nature but intrinsically are in interaction with omnipresent macroscopic
environments requires to go beyond weak coupling. It thus calls for an
urgent clarification of the thermodynamics of nanosystems and related 
issues. 

\ack  Financial support of the German Excellence Initiative via the
{\it Nanosystems Initiative Munich} (NIM) is gratefully
acknowledged.

\end{document}